# Tuning the generalized Hybrid Monte Carlo algorithm

A. D. Kennedy,[a*] Robert Edwards,[a] Hidetoshi Mino,[b] and Brian Pendleton[c],

[a]SCRI, Florida State University, Tallahassee, FL 32306–4052, USA

[b]Yamanashi University, Takeda 4, Kofu, 400 Japan

[c]Department of Physics & Astronomy, The University of Edinburgh,
The King's Buildings, Edinburgh EH9 3JZ, Scotland

We discuss the analytic computation of autocorrelation functions for the generalized Hybrid Monte Carlo algorithm applied to free field theory and compare the results with numerical results for the $O(4)$ spin model in two dimensions. We explain how the dynamical critical exponent $z$ for some operators may be reduced from two to one by tuning the amount of randomness introduced by the updating procedure, and why critical slowing down is not a problem for other operators.

## 1. GENERALIZED HMC

The work reported here extends the results first presented in [1]. We begin by recalling that a Markov Process will converge to some distribution of configurations if it is constructed out of update steps each of which has the desired distribution as a fixed point, and which taken together are ergodic. The generalized HMC algorithm is constructed out of two such steps.

### 1.1. Molecular Dynamics Monte Carlo

This consists of three parts: (1) *MD:* an approximate integration of Hamilton's equations on phase space which is exactly area-preserving and reversible; $U(\tau) : (\phi, \pi) \mapsto (\phi', \pi')$ where $\det U = 1$ and $U(\tau) = U(-\tau)^{-1}$. (2) A momentum flip $F : \pi \mapsto -\pi$. (3) *MC:* a Metropolis accept/reject test.

### 1.2. Partial Momentum Refreshment

This mixes the Gaussian-distributed momenta $\pi$ with Gaussian noise $\xi$:

$$\begin{pmatrix} \pi' \\ \xi' \end{pmatrix} = \begin{pmatrix} \cos\theta & \sin\theta \\ -\sin\theta & \cos\theta \end{pmatrix} \cdot F \begin{pmatrix} \pi \\ \xi \end{pmatrix}$$

The HMC algorithm is the special case where $\theta = \frac{\pi}{2}$. $\theta = 0$ corresponds to an exact version of the MD or microcanonical algorithm (which is in general non-ergodic). The L2MC algorithm of Horowitz [2,3] corresponds to choosing arbitrary $\theta$ but MDMC trajectories of a single leapfrog integration step.

## 2. LEAPFROG EVOLUTION

Consider a system of harmonic oscillators $\{\phi_p\}$ for $p \in \mathbb{Z}_V$. The Hamiltonian on phase space is $H = \frac{1}{2}\sum_{p\in\mathbb{Z}_V}\left(\pi_p^2 + \omega_p^2\phi^2\right)$. As the Hamiltonian is diagonal we shall temporarily suppress the index $p$ and set $\omega_p = 1$.

### 2.1. Evolution operator

The leapfrog discretization of Hamilton's equations can be written as

$$U(\delta\tau) = \begin{pmatrix} 1 - \frac{1}{2}\delta\tau^2 & \delta\tau \\ -\delta\tau + \frac{1}{4}\delta\tau^3 & 1 - \frac{1}{2}\delta\tau^2 \end{pmatrix}.$$

The most general area-preserving reversible linear mapping may be parameterized as

$$\begin{pmatrix} \cos[\kappa(\delta\tau)\delta\tau] & \dfrac{\sin[\kappa(\delta\tau)\delta\tau]}{\rho(\delta\tau)} \\ -\rho(\delta\tau)\sin[\kappa(\delta\tau)\delta\tau] & \cos[\kappa(\delta\tau)\delta\tau] \end{pmatrix}$$

where $\kappa$ and $\rho$ are even functions of $\delta\tau$. For leapfrog we find that $\kappa = 1 + \frac{1}{24}\delta\tau^2 + \frac{3}{640}\delta\tau^4 + O(\delta\tau^6)$ and $\rho = 1 - \frac{1}{8}\delta\tau^2 - \frac{1}{128}\delta\tau^4 + O(\delta\tau^6)$. With this parameterization it is easy to see that

*Talk presented by the first author. This work was partly supported by the DOE under grants #DE–FG05–85ER250000 and #DE–FG05–92ER40742.

the evolution operator for a trajectory of $\tau/\delta\tau$ leapfrog steps is

$$U(\tau) = \begin{pmatrix} \cos[\kappa(\delta\tau)\tau] & \dfrac{\sin[\kappa(\delta\tau)\tau]}{\rho(\delta\tau)} \\ -\rho(\delta\tau)\sin[\kappa(\delta\tau)\tau] & \cos[\kappa(\delta\tau)\tau] \end{pmatrix}.$$

## 3. ACCEPTANCE RATES

Given the evolution matrix we can compute the average Metropolis acceptance rate $\langle P_{\text{acc}}\rangle$. The probability distribution of $\delta H$ may be evaluated using Laplace's method to give an asymptotic expansion in the lattice volume $V$

$$P_{\delta H}(\xi) \sim \frac{1}{\sqrt{4\pi\langle\delta H\rangle}} \exp\left[-\frac{(\xi - \langle\delta H\rangle)^2}{4\langle\delta H\rangle}\right],$$

just as we would expect from the central limit theorem. The average Metropolis acceptance rate is thus

$$\langle P_{\text{acc}}\rangle \sim \operatorname{erfc}\left(\frac{1}{2}\sqrt{\langle\delta H\rangle}\right) = \operatorname{erfc}\left(\sqrt{\frac{1}{8}\langle\delta H^2\rangle}\right).$$

The acceptance rate is a function of the "scaling" variable $x \equiv V\delta\tau^4$: The explicit dependence is $\langle\delta H\rangle = \frac{1}{32} \times \frac{x}{V} \sum_{p\in\mathbb{Z}_V} (\sin\omega_p\tau)^2\omega_p^4$.

## 4. AUTOCORRELATION FUNCTIONS

### 4.1. Markov Processes

Let $(\phi_1, \phi_2, \ldots, \phi_N)$ be a sequence of field configurations generated by an equilibrated ergodic Markov process, and let $\langle\Omega(\phi)\rangle$ denote the expectation value of some connected operator $\Omega$. We may define an *unbiased estimator* $\bar\Omega$ over the finite sequence of configurations by $\bar\Omega \equiv \frac{1}{N}\sum_{t=1}^{N}\Omega(\phi_t)$, As usual, we define $C_\Omega(\ell) \equiv \frac{\langle\Omega(\phi_{t+\ell})\Omega(\phi_t)\rangle}{\langle\Omega(\phi)^2\rangle}$ as the *autocorrelation function* for $\Omega$. The variance of the estimator $\bar\Omega$ is

$$\langle\bar\Omega^2\rangle = \{1 + 2A_\Omega\}\frac{\langle\Omega(\phi)^2\rangle}{N}\left[1 + O\left(\frac{N_{\text{exp}}}{N}\right)\right],$$

where $A_\Omega \equiv \sum_{\ell=1}^{\infty} C_\Omega(\ell)$ is the *integrated autocorrelation function* for the operator $\Omega$ and $N_{\text{exp}}$ is the *exponential autocorrelation time.* This result tells us that on average $1 + 2A_\Omega$ correlated measurements are needed to reduce the variance by the same amount as a single truly independent measurement.

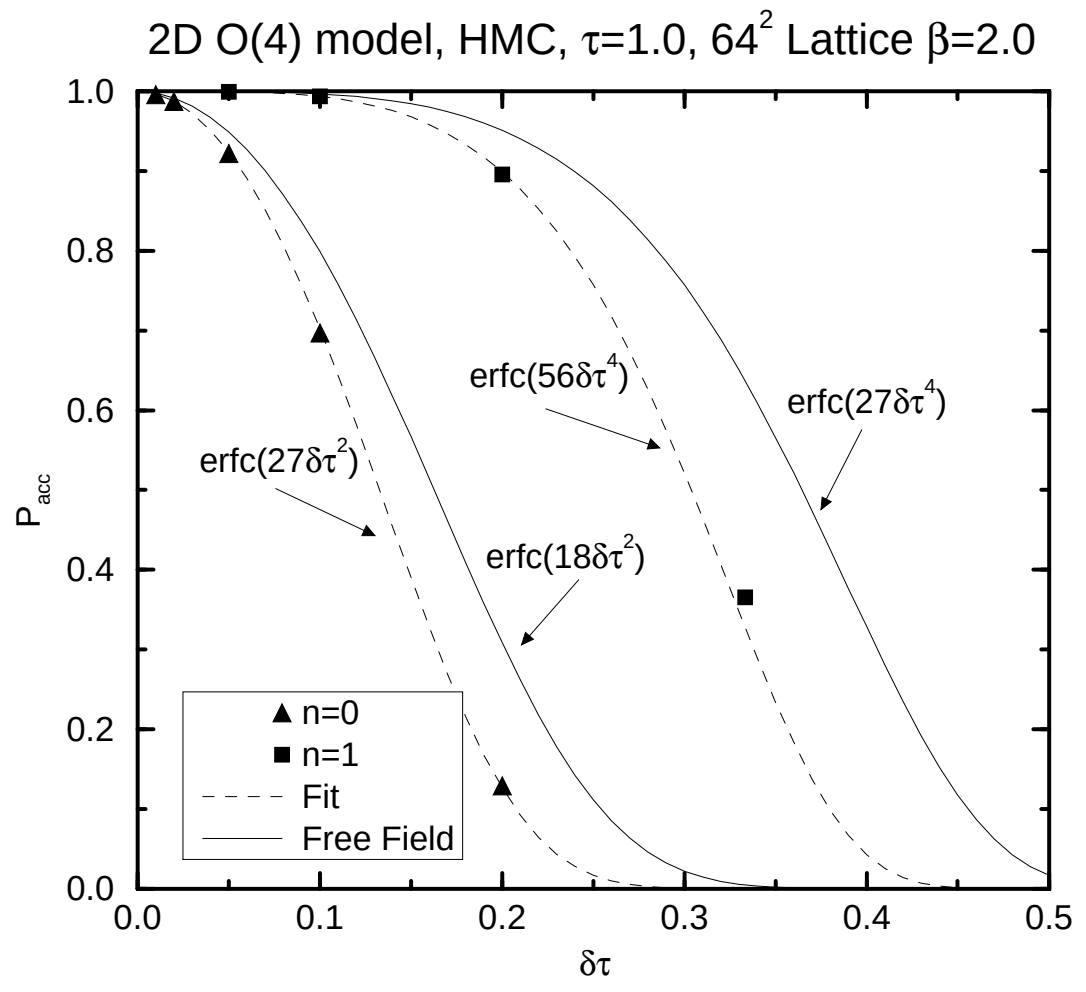

Figure 1. Comparison of acceptance rates for 2D $O(4)$ model with 2D free field theory; $n$ is the order of the integration scheme, with $n = 0$ for leapfrog. The higher order integration schemes have a scaling variable of the form $x = V\delta\tau^{4n+4}$.

### 4.2. Cost

We shall calculate $A_\Omega$ as a function of MD time. To a good approximation the cost $\mathfrak{C}$ of the computation is proportional to the total MD time for which we have to integrate Hamilton's equations. The cost per independent configuration is then $\mathfrak{C} \propto \frac{\tau}{\delta\tau}(1 + 2A_\Omega)$. The optimal trajectory length is obtained by minimizing the cost as a function of the parameters $\delta\tau$, $\tau$, and $\theta$ of the algorithm.

### 4.3. Autocorrelation functions for polynomial operators

We need to make a few simplifying assumptions: (1) The acceptance probability $P_{\text{acc}}$ for each trajectory may be replaced by its average value; we neglect correlations in the acceptance probability between successive trajectories. Including such correlations leads to seemingly intractable complications. It is not obvious that our assumption corresponds to any systematic approximation except, of course, that it is valid when $P_{\text{acc}} = 1$. (2) $\langle P_{\text{acc}}\rangle$ is assumed to be independent of trajectory length. This assumption is made purely for simplicity, as otherwise our results are expressed in terms of particularly disgusting integrals.

We may ignore the corrections of non-leading order in $\delta\tau$ to the MD evolution operator because for any given value of $\langle P_{\rm acc}\rangle$ there is a corresponding value of $x$ which is $O(1)$, and thus $\delta\tau$ is of order $V^{-1/4}$. These corrections therefore only contribute to the autocorrelations through the acceptance rate itself.

We shall choose each trajectory $\tau$ length independently from some distribution $P_R(\tau)$, as this avoids the lack of ergodicity caused by choosing a fixed trajectory length which is a rational multiple of the period of any mode of the system. This is a disease of free field theory which in interacting models is removed to some extent by mode coupling.

The integrated autocorrelation function for the connected squared magnetization $M^2$ with exponentially distributed trajectory lengths, $P_R(t) = e^{-t/\tau}$, is

$$\frac{\begin{array}{c}-4P_{\rm acc}^2\omega^2\tau^2\cos^3\theta + 8P_{\rm acc}\omega^2\tau^2\cos^3\theta - \\ -4\omega^2\tau^2\cos^3\theta + 2P_{\rm acc}\cos^3\theta - \cos^3\theta - \\ -4P_{\rm acc}\omega^2\tau^2\cos^2\theta - 4\omega^2\tau^2\cos^2\theta - \cos^2\theta - \\ -2P_{\rm acc}^2\omega^2\tau^2\cos\theta - 6P_{\rm acc}\omega^2\tau^2\cos\theta + \\ +4\omega^2\tau^2\cos\theta - 2P_{\rm acc}\cos\theta - \cos\theta - \\ -2P_{\rm acc}\omega^2\tau^2 - 4\omega^2\tau^2 - 1\end{array}}{2P_{\rm acc}\omega^2\tau^2\left(\cos^2\theta - 1\right)\left(P_{\rm acc}\cos\theta - \cos\theta - 1\right)}$$

The integrated autocorrelation function for $M^2$ with fixed length trajectories, $P_R(t) = \delta(t-\tau)$, is

$$\frac{\begin{array}{c}-2P_{\rm acc}\cos\theta\sin^2\theta + \cos\theta\sin^2\theta - \\ -2P_{\rm acc}\sin^2(\omega\tau)\sin^2\theta + \sin^2\theta + \\ +P_{\rm acc}\sin^2(\omega\tau)\cos\theta + P_{\rm acc}\sin^2(\omega\tau)\end{array}}{P_{\rm acc}\sin^2(\omega\tau)\left(\cos\theta + 1\right)\sin^2\theta}$$

## 5. RESULTS FOR $C_{M^2}(T)$ AND $A_{M^2}$

In order to understand the results let us consider the special case where $\theta = 0$ (HMC) and $P_{\rm acc} = 1$. In this case the Laplace transform of the autocorrelation function is

$$F_{M^2}(\beta) = \frac{r^3 + 2r^2\beta + r\beta^2 + 2m^2 r}{\beta^3 + 2r\beta^2 + (r^2 + 4m^2)\beta + 2m^2 r},$$

where we have written $r = 1/\tau$ and $\omega = m$.

### 5.1. Autocorrelation functions

The fact that $F_{M^2}(\beta)$ is a rational function in $\beta$ tells us that the autocorrelation function is a sum of exponentials. The exponents are the roots of the cubic denominator, and they are all real or one is real and the other two are complex conjugates depending on the value of the mean trajectory length $1/r$.

### 5.2. Integrated autocorrelation function

This is just $A_{M^2} = F_{M^2}(0) = 1 + \frac{1}{2}\left(\frac{1}{m\tau}\right)^2$. Minimizing the cost by choosing $\tau_{\rm opt}$ such that $\left.\frac{d\mathfrak{C}}{d\tau}\right|_{\tau=\tau_{\rm opt}} = 0$ we obtain $\tau_{\rm opt} = 1/\sqrt{3}m$ and thence $A_{M^2}(\tau_{\rm opt}) = 5/2$. The optimum trajectory length depends upon the operator being considered.

### 5.3. Dynamical critical exponent

One of the most relevant measures of the effectiveness of an algorithm for studying continuum physics is the exponent $z$ relating the cost $\mathfrak{C}$ to the correlation length $\xi$ of the system. For free field theory the correlation length is just the inverse mass, and thus we have $z = 2$ if we hold $\tau$ constant, but $z = 1$ if we choose $\tau = \tau_{\rm opt}$ which grows as $1/m$.

### 5.4. Optimal choice for $\theta$

For the generalized HMC algorithm we should minimize the cost by varying both $\tau$ and $\theta$. The optimal choice of parameters when $P_{\rm acc} = 1$ is to take $\theta \to 0$ and $\tau \to 0$, but this ignores the fact that the cost does not decrease when we take $\tau$ smaller than the $\delta\tau$ required to obtain a reasonable Metropolis acceptance rate. If we choose $\tau_{\rm opt} = \delta\tau$ (L2MC) and the corresponding value for $\theta_{\rm opt}$ we find that the cost is less than for the HMC case, but only by a constant factor. As the cost is only defined up to an implementation dependent constant factor anyhow we may conclude that generalized HMC does not appear to promise great improvements over HMC.

### 5.5. L2MC

Horowitz suggested that the L2MC algorithm was better than HMC not for the reasons discussed above, but because the acceptance rate is much higher for fixed $\delta\tau$. Unfortunately, when $\theta \neq \frac{\pi}{2}$ we must flip the momenta upon rejecting a tra-jectory, and this causes the system to perform a random walk unless the acceptance rate is very close to unity. Our data shows that in practice the cheapest solution appears to be HMC.

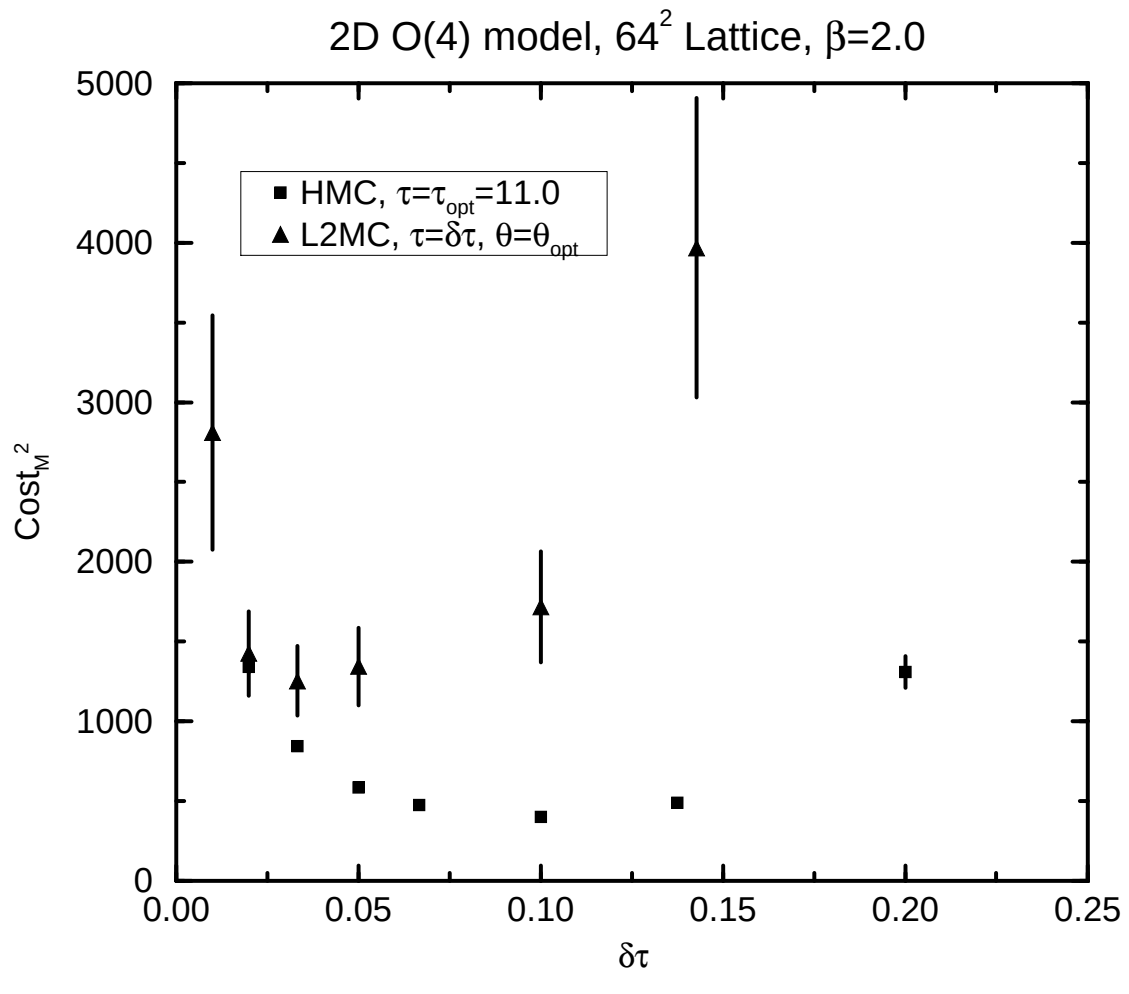

Figure 2. Comparison of the HMC and L2MC algorithms.

## 6. RESULTS FOR $A_E$

So far all our results have been for the operator $M^2$, but this is a rather special case as it couples solely to the slowest mode of the system. Let us now investigate the properties of the energy operator $E = \frac{1}{2}\sum_p \omega_p^2 \phi_p^2$.
The Laplace transform of the connected autocorrelation function for the energy is $F_E(\beta) = \frac{1}{V}\sum_p \frac{r^3+2r^2\beta+r\beta^2+2\omega_p^2 r}{\beta^3+2r\beta^2+(r^2+4\omega_p^2)\beta+2\omega_p^2 r}$. The integrated autocorrelation function for the energy is

$$A_E = F_E(0) = 1 + \frac{1}{2\tau^2}\sum_p \frac{1}{\omega_p^2} \equiv 1 + \frac{1}{2\tau^2}\sigma_{m^2}^{(-1)},$$

and the optimal trajectory length is $\tau_{\rm opt} = \sqrt{\sigma_{m^2}^{(-1)}/3}$, leading to an integrated autocorrelation function value of $A_E(\tau_{\rm opt}) = 5/2$.
In order to determine the dynamical critical exponent $z$ we need to evaluate the spectral sum $\sigma_{m^2}^{(-1)}$.

### 6.1. Spectral sums for 2D free field theory

Using Poisson resummation we find that in the thermodynamic limit

$$\frac{1}{V}\sum_{p_x,p_y} \omega_p^{-2} = \frac{2\sqrt{ab}}{\pi(1-ab)} K\left(\frac{b-a}{1-ab}\right),$$

where $a = \frac{1}{2}m^2 + 3 - \sqrt{(\frac{1}{2}m^2+3)^2 - 1}$, $b = \frac{1}{2}m^2 + 1 - \sqrt{(\frac{1}{2}m^2+1)^2-1}$, and $K(k)$ is a complete elliptic integral. For small $m$ we find that $\sigma_{m^2}^{(-1)} \approx \frac{1}{4\pi}\ln\frac{32}{m^2}$, and hence $z = 0$. Note that this does not mean that the cost does not increase with increasing $\xi$, but that it increases only logarithmically.

## 7. CONCLUSIONS

(1) Too little noise increases critical slowing down because the system is too weakly ergodic. (2) Too much noise increases critical slowing down because the system takes a drunkard's walk through phase space. (3) To attain $z = 1$ for all operators (and especially for the exponential autocorrelation time) one must be able to tune the amount of noise suitably. (4) For operators such as $E$, critical slowing down is unimportant because they only couple weakly to the slowest modes of the system.

## REFERENCES

1. A. D. Kennedy and B. J. Pendleton. Acceptances and autocorrelations in Hybrid Monte Carlo. In Urs M. Heller, A. D. Kennedy, and Sergiu Sanielevici, editors, *Lattice '90*, volume B20 of *Nuclear Physics (Proceedings Supplements)*, pages 118–121, 1991. Talk presented at "Lattice '90," Tallahassee.
2. Alan Horowitz. The second order Langevin equation and numerical simulations. *Nucl. Phys.*, B280[FS18](3):510–522, 1987.
3. Alan Horowitz. A generalized guided Monte Carlo algorithm. *Phys. Lett.*, B268:247–252, 1991.